\documentclass[aps,prl,reprint,twocolumn,amsmath,amssymb,showpacs,superscriptaddress]{revtex4-1}

\usepackage{color}
\usepackage{textcomp} 
\usepackage{mathrsfs,amsmath}
\usepackage{graphicx}
\usepackage{dcolumn}
\usepackage[mathlines]{lineno}
\usepackage{hyperref}
\usepackage{xcolor}
\hypersetup{
    colorlinks,
    linkcolor={red!50!black},
    citecolor={blue!50!black},
    urlcolor={blue!80!black}
}
\usepackage{bm}
\usepackage{epstopdf}
\usepackage{soul}
\usepackage{epsfig}
\usepackage{bbold}
\usepackage{braket}
\usepackage{physics}
\usepackage{gensymb}
\usepackage{amsmath,amssymb}

\begin{document}

\title{Topological-charge-dependent dichroism and birefringence of optical vortices}% Force line breaks with \\
\author{Kayn A. Forbes}
\email{K.Forbes@uea.ac.uk}

\affiliation{School of Chemistry, University of East Anglia, Norwich Research Park, Norwich NR4 7TJ, United Kingdom}

\author{Dale Green}

\affiliation{Physics, Faculty of Science, University of East Anglia, Norwich Research Park, Norwich NR4 7TJ, United Kingdom}

\begin{abstract}
Material anisotropy and chirality produce polarization-dependent light-matter interactions. Absorption leads to linear and circular dichroism, whereas elastic forward scattering produces linear and circular birefringence. Here we highlight a form of dichroism and birefringence whereby ordered generic media display locally different absorption and scattering of a focused vortex beam  that depends upon the sign of the topological charge $\ell$. The light-matter interactions described in this work manifest  purely through dominant electric-dipole coupling mechanisms and depend on the paraxial parameter to first-order. Previous topological-charge-dependent light-matter interactions required the significantly weaker higher-order multipole moments and are proportional to the paraxial parameter to second-order. The result represents a method of probing the nano-optics of advanced materials and the topological properties of structured light.  
\end{abstract}

\maketitle

\section{Introduction}

Light-matter interactions in oriented and anisotropic media depend strongly on the relative orientation of the polarization vector of light with respect to the charge and current displacements. Optical anisotropy, for example, is responsible for linear birefringence (elastic forward scattering) and dichroism (absorption) of light \cite{barron2009molecular}. The physical origin of these phenomena is intuitive: in anisotropic materials the ability for the electric field of light to cause oscillations in charge and current distributions strongly depends on the orientation of the material with respect to the polarization state of the incident electromagnetic field. Optical anisotropy has long been exploited to produce optical elements that transform the degrees of freedom of light, e.g. waveplates \cite{saleh2019fundamentals}. More recent applications include engendering spin-orbit interactions of light \cite{bliokh2015spin} in liquid crystals and metamaterials. Distinct, but also polarization-dependent is optical activity \cite{barron2009molecular}: Chiral materials exhibit polarization-dependent absorption and refraction of circularly polarized light (CPL) through circular dichroism and circular birefringence (optical rotation), respectively.

Light-matter interactions in anisotropic and ordered materials are generally subject to input light propagating in the paraxial regime: the electric field is polarized transverse (\emph{xy} plane) to the direction of propagation (\emph{z}) of the beam. This form of light is well-described by the standard Stokes vector, being polarized in two-dimensions (2D polarized light). However, due to the finite spatial confinement of light sources, all electromagnetic fields possess polarized components in the direction of propagation: longitudinal fields. These longitudinal field components (also referred to as non-paraxial), which make the light polarized in three-dimensions (3D) \cite{alonso2023geometric}, are responsible for a remarkable number of extraordinary properties of light in nano-optics \cite{bliokh2015spin,chekhova2021polarization,novotny2012principles}. Crucially, the magnitude of longitudinal components is dependent on the degree of spatial confinement of the field. Strongly confined fields - evanescent waves and tightly focused laser beams, for example - possess longitudinal components that can produce interactions with materials that have observable consequences, whereas free space plane waves and well-collimated beams do not. 

Assuming propagation along \emph{z}, the longitudinal \emph{z}-polarized electric and magnetic fields of spatially confined light are in general $\pi/2$  out-of-phase with their respective transverse \emph{xy}-polarized field components. This leads to a cyclic rotation (or spinning) of the electromagnetic field transverse to the direction of propagation, producing transverse spin angular momentum (SAM) of light and the so-called photonic wheels \cite{shi2021spin,bliokh2015transverse,aiello2015transverse,eismann2021transverse}, for example.  Optical vortices are orbital angular momentum (OAM) carrying modes of light due to their azimuthal phase $\text{e}^{i\ell\phi}$, where $\ell \in \mathbb{Z}$, leading to an OAM of $\ell\hbar$  per photon. The sign of $\ell$  determines the handedness (geometrical chirality) of the twisted wavefront of vortex beams: $\ell>0$  are left-handed, $\ell<0$  are right-handed. Beyond established widespread applications \cite{forbes2021structured,shen2019optical, nape2023quantum}, optical vortices have recently been shown to exhibit extraordinary optical activity and optical chirality properties due to their longitudinal electromagnetic fields \cite{green2023optical,forbes2023customized,forbes2022optical}. Optical vortex modes, in addition to the ubiquitous $\pi/2$ out-of-phase longitudinal component of electromagnetic fields, possess an additional \emph{in-phase} longitudinal component which is significantly strengthened due to the topological properties of the beam. Here we highlight that this in-phase longitudinal field component leads to the manifestation of a local dichroism and birefringence of focused vortex light which depends on \emph{the sign of the topological charge $\ell$}: {topological-charge-dependent} dichroism and birefringence.

\section{Topological-charge-dependent polarization}

The electric field for an arbitrarily polarized \emph{z}-propagating monochromatic Laguerre-Gaussian (LG) beam in cylindrical coordinates (\emph{r}, $\phi$, \emph{z})  which includes terms up to first order in the paraxial parameter $1/kw$  (we constrain ourselves to this approximation throughout the manuscript), where $k$  is the wavenumber and $w$ the beam waist, is \cite{supplement}

\begin{align}
\mathbf{E} &= \bigl[\alpha\mathbf{\hat{x}}
+\beta\mathbf{\hat{y}} 
+\frac{i}{k}\mathbf{\hat{z}}
\bigl\{\alpha\bigl(\gamma\cos\phi-\frac{i\ell}{r}\sin\phi\bigr)
\nonumber\\ &+\beta\bigl(\gamma\sin\phi+\frac{i\ell}{r}\cos\phi\bigr)\bigr\} \bigr]u_{\ell,p}^{\text{LG}}(r,\phi,z),
\label{eq:1}
\end{align}
where $\gamma = \frac{\abs{\ell}}{r}-\frac{2r}{w^2}-\frac{4rL_{p-1}^{\abs{\ell}+1}}{w^2L_{p}^{\abs{\ell}}}$;  $L_{p}^\abs{\ell}$ is the generalized Laguerre polynomial; $\alpha$, $\beta$ are the Jones vector coefficients: $\abs{\alpha}^2+\abs{\beta}^2=1$  ; $\ell \in \mathbb{Z}$ and   $p \in \mathbb{Z^+}$  are the topological charge and radial index, respectively; $u_{\ell,p}^{\text{LG}}(r,\phi,z)$ is the well-known amplitude distribution for LG beams (see \cite{supplement}), which includes the all-important azimuthal phase $\text{e}^{i\ell\phi}$. 

The \emph{x} and \emph{y} components of Eq. \eqref{eq:1} are the transverse electric field (with respect to the direction of propagation) and account for the 2D polarization state, described using the standard theory of paraxial optics and a 2x2 polarization matrix \cite{chekhova2021polarization}. The \emph{z} component is the longitudinal field. The full field, including the non-paraxial longitudinal part, is referred to as 3D polarized and requires a 3x3 polarization matrix to be described \cite{alonso2023geometric}. 

\begin{figure} [ht]
\includegraphics[]{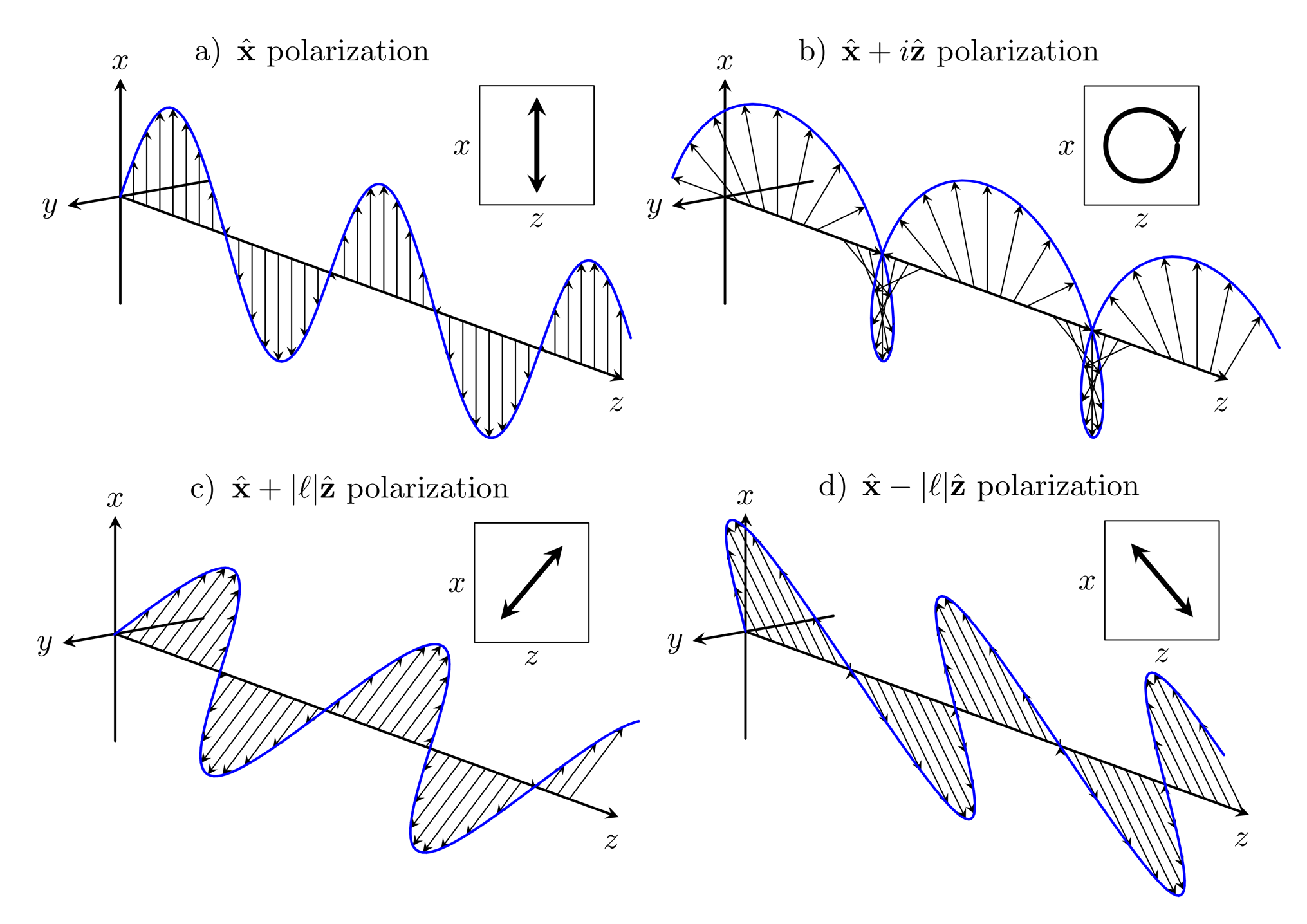}
\caption{a) 2D \emph{x}-polarized electric field vector for a \emph{z}-propagating beam e$^{ikz}$; b) 2D \emph{x}-polarized light with ${\pi}/{2}$ out-of-phase longitudinal component leading to an elliptical polarization vector in the \emph{xz}-plane. It is this spinning transverse electric field vector which is responsible for the transverse spin angular momentum of light; c) 2D \emph{x}-polarized optical vortex light with an in-phase longitudinal field component for positive values of $\ell$. The polarization vector is tilted in the positive \emph{z}-direction in the \emph{xz}-plane; d) same as c but for negative values of $\ell$. In this case the polarization vector is tilted towards the negative \emph{z}-direction. As is clear from Eq. \eqref{eq:1}, analogous results would manifest for other 2D states of polarization.}
    \label{fig:1}
\end{figure}

To most clearly elucidate the topological-charge-dependent polarization we concentrate on 2D linearly polarized light in this Section ($\alpha$  and $\beta$  are real). Note, however, that 2D circularly polarized light ($\beta=i\sigma/\sqrt{2}$) contributes a small in-phase transverse-longitudinal contribution proportional to the helicity $\sigma$, but is not topologically dependent \cite{supplement}. The imaginary longitudinal field terms in Eq. \eqref{eq:1} dependent on $\gamma$ are $\frac{\pi}{2}$  out of phase with the transverse components  due to the \emph{i} prefactor of the \emph{z}-components (as discussed in the Introduction). In calculating the (electric) cycle-averaged spin angular momentum density of the field Eq. \eqref{eq:1} using $\overline{\mathbf{s}}_\text{E}=\text{Im}(\mathbf{E^*}\times{{\mathbf{E}}})$ it is simple to show the imaginary longitudinal components lead to a non-zero transverse spin \cite{bliokh2015transverse}. However, the real longitudinal terms in Eq. \eqref{eq:1}, which depend on $\ell$ , are in-phase with the transverse components. Being dependent on $\ell$  means that this phenomenon of in-phase longitudinal and transverse field components for real $\alpha$ and $\beta$ is unique to optical vortex modes, and does not manifest in fields where $\ell = 0$, e.g. Gaussian beams or evanescent waves (note this is not the case for 2D polarization with non-zero helicity, see \cite{supplement}).  Crucially, the sign of $\ell$ determines the orientation of the polarization state in \emph{xz} (see Fig. \ref{fig:1}) or \emph{yz} planes. It is important to note an analogous behaviour is observed for the magnetic field, however due to the electric-biased nature of most dielectric materials the optical magnetic light-matter interaction is significantly weaker than the electric-dipole coupling we study herein. 

This property of $\ell$ dependent 3D polarization orientation for vortex modes means, in a rather generic sense, that any observable of a light-matter interaction which depends on the polarization vector of the beam (with respect to the material orientation) is modified by the topological structure of generic vortex modes (LG, Bessel, etc.). For example, the most fundamental mechanisms of absorption and scattering of optical vortex light by materials will exhibit this $\ell$ dependence under suitable circumstances. We now determine the conditions this $\ell$ dependent orientation of the polarization state influences absorption and forward elastic scattering of these beams.

\section{Topological-charge-dependent dichroism (TD)}

The absorption of light by matter to leading order is described by the interaction Hamiltonian truncated to electric-dipole approximation: $H_\text{int}=-\mu_i{E_i}$   (repeated subscript indices imply Einstein summation convention). The Fermi golden rule tells us the rate of absorption  $W_{I\rightarrow{F}}$, i.e. the rate of transfer of energy from the electromagnetic field to the system is $W_{I\rightarrow{F}}=\frac{2\pi}{\hbar}|\langle F \vert\;H_\text{int}\;\vert I \rangle|^2\rho(E_{FI})$ \cite{craig1998molecular}. The density of states $\rho(E_{FI})$ is specific to the given light-matter system. Using Eq. \eqref{eq:1}, the amplitude for absorption is:
\begin{align}
\langle F \vert\;H_\text{int}\;\vert I \rangle &=
\bigl[\alpha{\hat{x_i}}
+\beta{\hat{y_i}} +\frac{i}{k}{\hat{z_i}}
\bigl\{\alpha\bigl(\gamma\cos\phi-\frac{i\ell}{r}\sin\phi\bigr)
\nonumber\\ 
&+\beta\bigl(\gamma\sin\phi+\frac{i\ell}{r}\cos\phi\bigr)\bigr\}\bigr] \mu_i u_{\ell,p}^{\text{LG}},
\label{eq:2}
\end{align}
where  $\langle \psi_F \vert\;\mu_i\;\vert \psi_I \rangle \equiv \mu_i^{FI} \equiv \mu_i $ . Retaining terms up to order $\frac{1}{kw}$  (sometimes referred to as the paraxial approximation, though somewhat a misnomer as it includes the non-paraxial longitudinal field) we have for the absorption probability (after being time-averaged over one period of oscillation):
\begin{figure*}
    \includegraphics[]{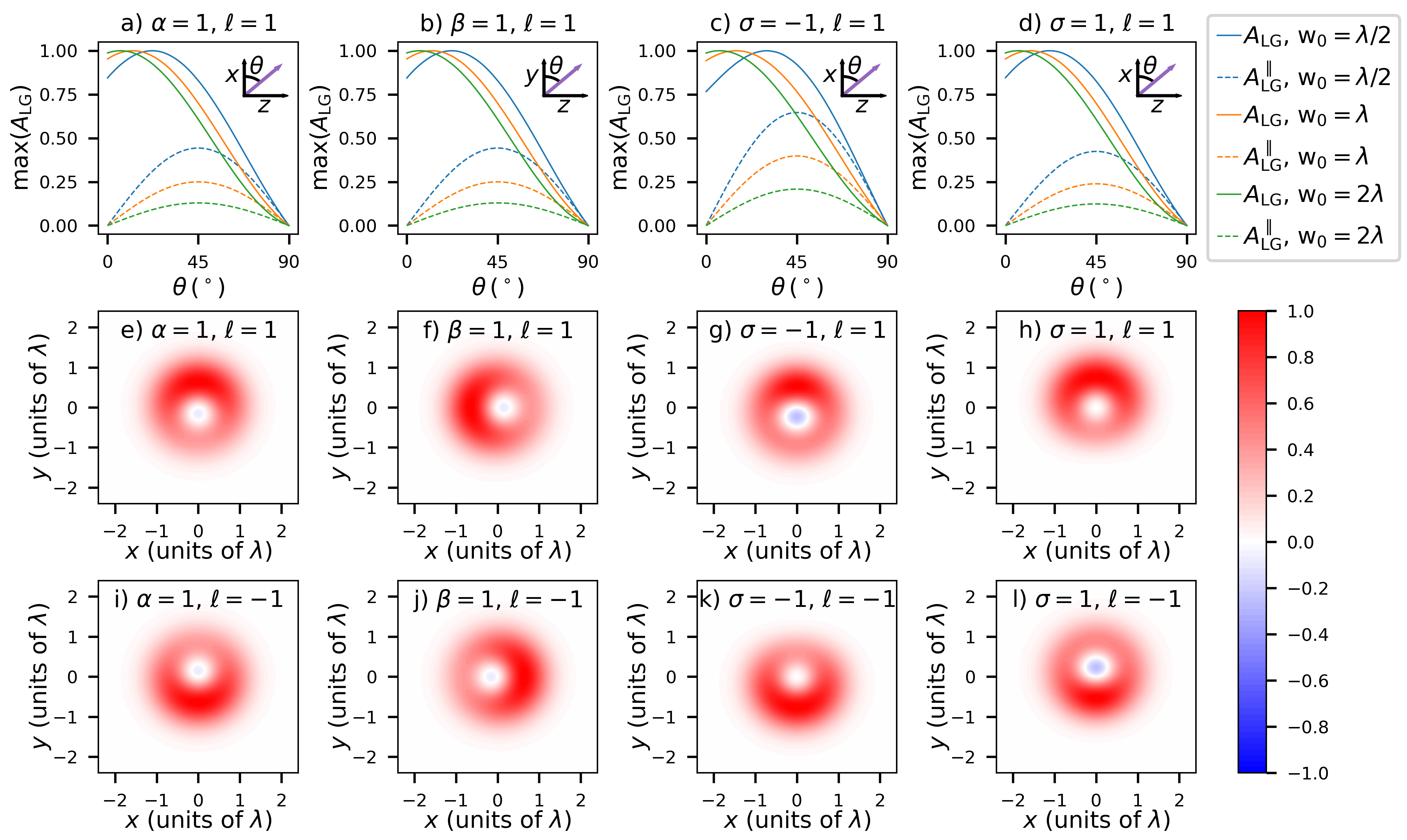}
    \caption{ a)--d) Dipole orientation dependence of Eq. \eqref{eq:3} for (solid lines) maximum total absorption, $A_\text{LG}$, individually normalised, (dashed lines) maximum TD contribution, $A_\text{LG}^\parallel$, normalized to the maximum of the solid line of the same color. Insets show dipole orientation with respect to $\theta$ and given axes. e)--l) Focal plane spatial distributions of Eq. \eqref{eq:3} with $\theta=45^\circ$, with dipole orientation matching the inset of the corresponding a)--d) of the same column. $p=0$ in all cases and the magnitude of all dipoles $|\mu_i| = 1$.}
    \label{fig:2}
\end{figure*}

\begin{align}
|\langle F \vert\;H_\text{int}\;\vert I \rangle|^2  &=
\mu_i\mu_j\bigl[|\alpha|^2{\hat{x_i}\hat{x_j}} + |\beta|^2{\hat{y_i}\hat{y_j}}+ 2\Re\alpha\beta^*\hat{x_i}\hat{y_j}
\nonumber\\ 
&+2\hat{x_i}\hat{z_j}\frac{1}{k}\bigl\{(|\alpha|^2\frac{\ell}{r}-\gamma\Im\alpha^*\beta)\sin\phi \nonumber \\
&-\frac{\ell}{r}\Re\alpha\beta^*\cos\phi\bigr\} 
+2\hat{y_i}\hat{z_j}\frac{1}{k}\bigl\{(-\gamma\Im\alpha\beta^* \nonumber \\
&- \frac{\ell}{r}|\beta|^2) \cos\phi +\frac{\ell}{r}\Re\alpha\beta^*\sin\phi \bigr\}\bigr]|u_{\ell,p}^{\text{LG}}|^2.
\label{eq:3}
\end{align}
The first three terms in Eq. \eqref{eq:3}, zeroth-order with respect to the paraxial parameter $1/kw$, are the well-known contributions to absorption of light under paraxial conditions \cite{barron2009molecular, craig1998molecular}; the remaining terms describe the influence that a focused vortex beam has on absorption. These terms manifest through the interference between the 2D polarized transverse field components and the first-order longitudinal component, and are thus proportional to the paraxial parameter to first order. 

There are two important limits of Eq. \eqref{eq:3} we can distinguish: whether the input beam is 2D linearly polarized (in which case $\alpha$ and $\beta$ are real); or whether the input beam is 2D circularly polarized: $\alpha=1/\sqrt{2}$, $\beta={i \sigma }/\sqrt{2}$, where the helicity is $\sigma = \pm 1 $, the upper-sign corresponding to left-handed CPL and the bottom right-handed CPL. For the 2D linear case Eq. \eqref{eq:3} becomes

\begin{align}
|\langle F \vert\;H_\text{int}\;\vert I \rangle|^2  &=
\mu_i\mu_j\bigl[|\alpha|^2{\hat{x_i}\hat{x_j}} + |\beta|^2{\hat{y_i}\hat{y_j}}+ 2\Re\alpha\beta^*\hat{x_i}\hat{y_j}
\nonumber\\ 
&+\frac{2\ell}{kr}\bigl(\hat{x_i}\hat{z_j}\bigl\{|\alpha|^2\sin\phi-\Re\alpha\beta^*\cos\phi\bigr\} \nonumber \\
&+\hat{y_i}\hat{z_j}\bigl\{\Re\alpha\beta^*\sin\phi -|\beta|^2 \cos\phi  \bigr\}\bigl)\bigr]|u_{\ell,p}^{\text{LG}}|^2.
\label{eq:4}
\end{align}
The spatial distribution of Eq. \eqref{eq:4} and dipole orientation dependence in the focal plane for 2D $x$-polarized and 2D $y$-polarized input beams are displayed in Fig. \ref{fig:2} a), e) i) and b), f), j), respectively. The azimuth of the 2D polarization state acts to rotate the spatial distribution of absorption, e.g. compare Fig. \ref{fig:2} e) to Fig. \ref{fig:2} f). More importantly, it is clear that the sign of the topological charge (wavefront handedness) leads to a differential absorption of the light by the material: $W_{I\rightarrow{F}}^{\ell} \neq W_{I\rightarrow{F}}^{-\ell} $. This is analogous to linear and circular dichroism, but the differential effect stems from the topological charge of the vortex beam: topological-charge-dependent dichroism. In the case of 2D CPL Eq. \eqref{eq:3} becomes

\begin{align}
|\langle F \vert\;H_\text{int}\;\vert I \rangle|^2  &=
\mu_i\mu_j\bigl[\frac{1}{2}{\hat{x_i}\hat{x_j}} + \frac{1}{2}{\hat{y_i}\hat{y_j}}
\nonumber\\ 
&+\hat{x_i}\hat{z_j}\frac{1}{k}\bigl\{\frac{\ell}{r}-\gamma\sigma \bigr\}\sin\phi \nonumber \\
&
+\hat{y_i}\hat{z_j}\frac{1}{k}\bigl\{\gamma\sigma 
- \frac{\ell}{r}   \bigr\}\cos\phi\bigr]|u_{\ell,p}^{\text{LG}}|^2.
\label{eq:5}
\end{align}
The spatial distribution of Eq. \eqref{eq:5} and dipole orientation dependence in the focal plane for 2D right and left circularly polarized input beams are displayed in Fig. \ref{fig:2} c), g), k) and d), h), l), respectively. We readily see an interplay of terms depending on $\sigma$ and $\ell$. Namely, the case of parallel SAM and OAM (sgn$\ell$=sgn$\sigma$) differs from the anti-parallell SAM and OAM (sgn$\ell$$\neq$sgn$\sigma$): the spatial distribution is more drastically altered by altering the wavefront handedness than the 2D circular polarization handedness, e.g. compare Fig. \ref{fig:2} g) and k) to g) and h). This is because the topological charge influences the 3D polarization state significantly more than the 2D polarization helicity \cite{supplement}. Furthermore, comparing Fig. \ref{fig:2} c) to Fig. \ref{fig:2} d)  highlights how the TD mechanism is significantly larger for the anti-parallel case compared to the parallel case: e.g. for $w_0=\lambda$ the ratio of the TD contribution to absorption $A_\text{LG}^\parallel$ versus the standard paraxial term is 84\%, whereas for the parallel case it is 50\%. This behaviour mirrors that which is known for properties (e.g. intensity) of vortex beams due to longitudinal field components \cite{bliokh2015spin}. 
 
It is important to appreciate the significant magnitude of the TD mechanism. The intensity of a tightly focused vortex beam, proportional to $\mathbf{E}\cdot\mathbf{E}^*$, consists of the inner product of the dominant 2D transverse fields producing a contribution which is zeroth-order in the paraxial parameter and the inner product of the first-order longitudinal components which yield a contribution that is second-order, i.e. $\propto 1/(kw)^2$. Compared to the 84\% ratio for anti-parallel discussed above, the second-order contribution to the intensity proportional to $1/(kw)^2$ is $\approx$ 10\% in optimal conditions for $w_0=\lambda$ relative to the zeroth-order fields \cite{forbes2021relevance}  (though readily observed \cite{iketaki2007investigation}). This highlights the significantly larger TD effect, proportional to first-order in the paraxial parameter $1/(kw)$.

TD does not manifest in isotropic media (fluids and gases). This is readily seen by rotational averaging Eq. \eqref{eq:3} using the well-known second-rank tensor average: e.g. $\hat{x_i}\hat{z_j}\langle{\mu_i}{\mu_j} \rangle=\delta_{ij}\hat{x_i}\hat{z_j}|\mu|^2/3=0$. One important application of TD would therefore be as a sensitive measure of partial or local order. 

\begin{figure*}
    \includegraphics[]{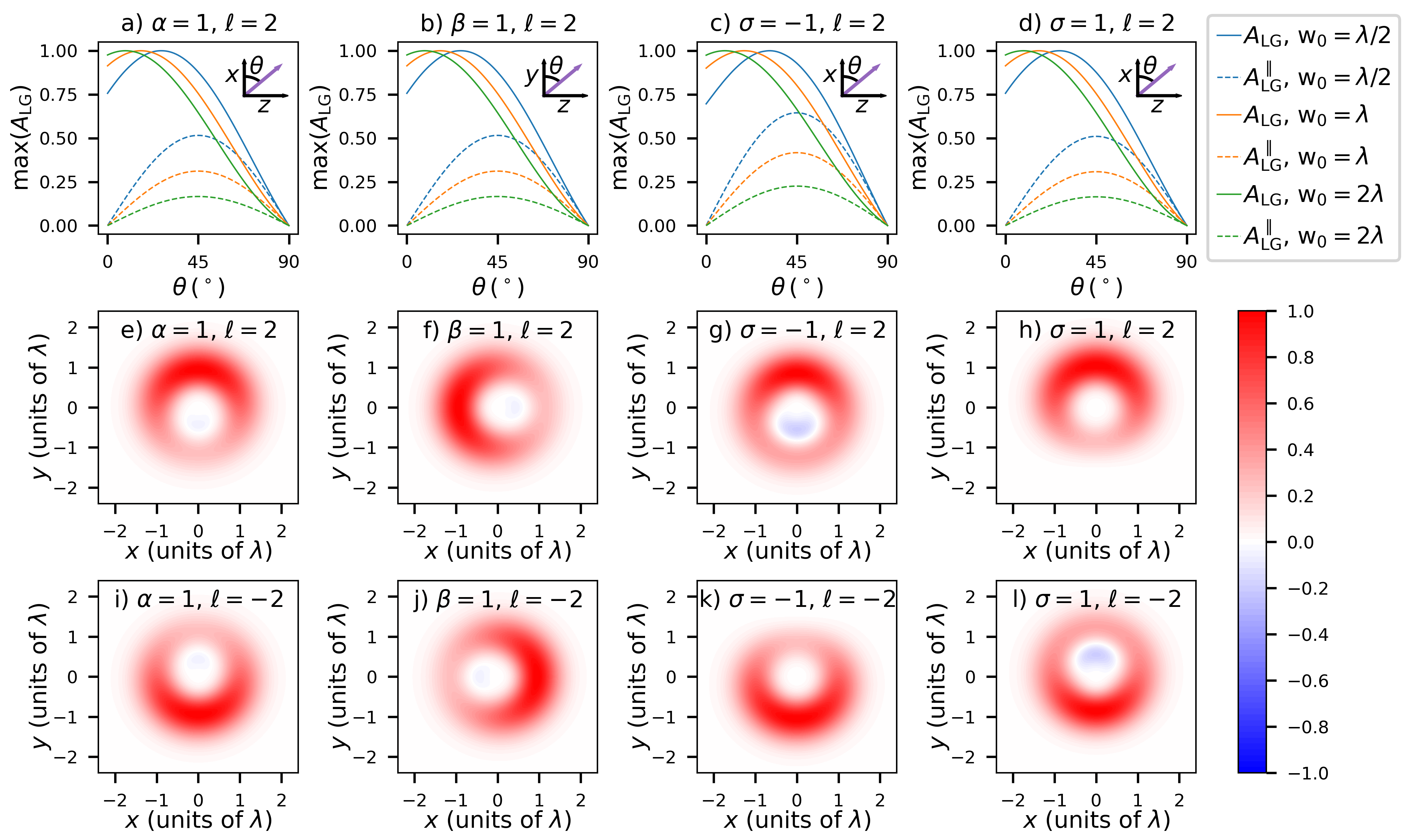}
    \caption{Same as Fig.~2 but for $\ell=2$.}
    \label{fig:3}
\end{figure*}

\begin{figure*}
    \includegraphics[]{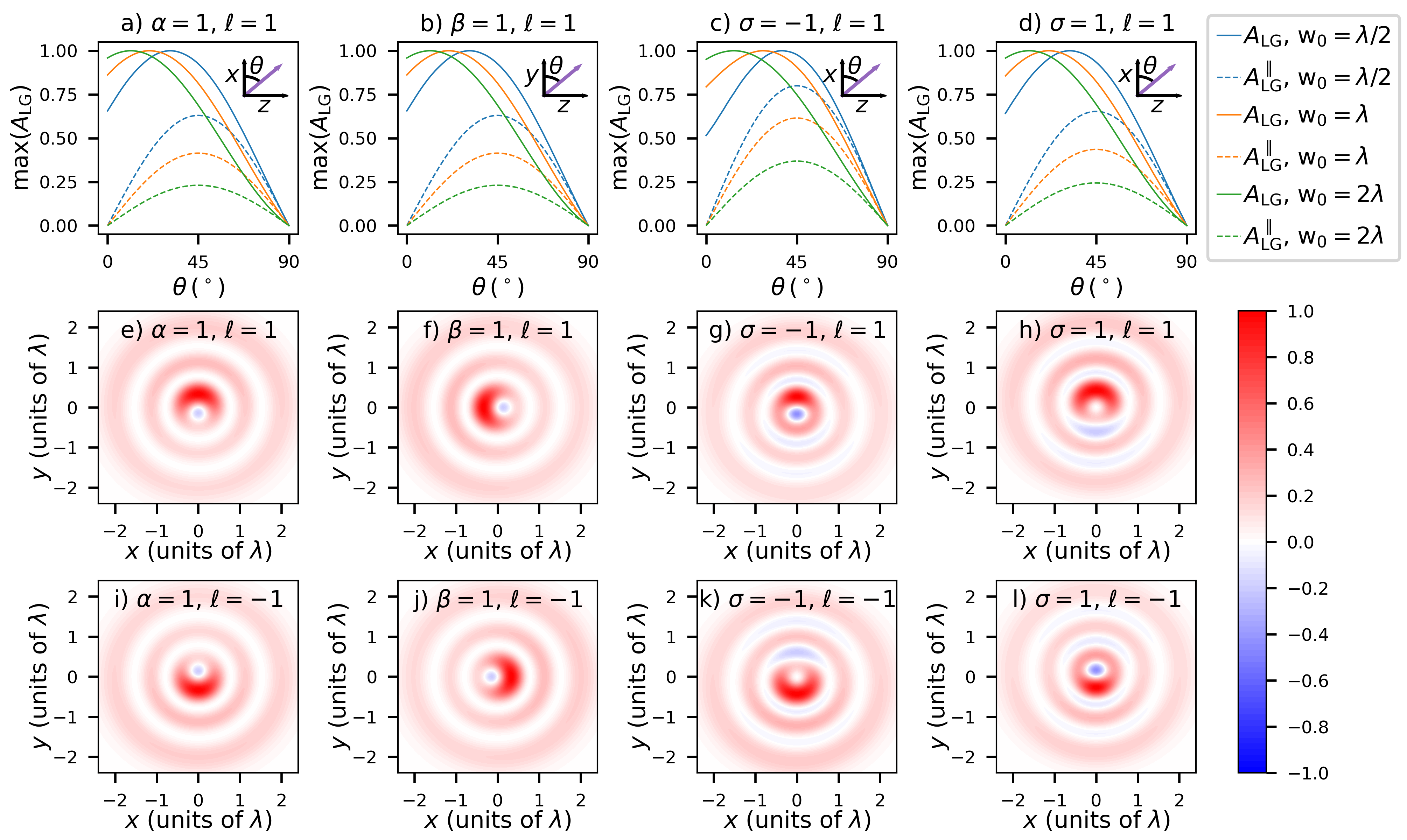}
    \caption{Same as Fig.~2 but for $p=2$.}
    \label{fig:4}
\end{figure*}

\begin{figure*}
    \includegraphics[]{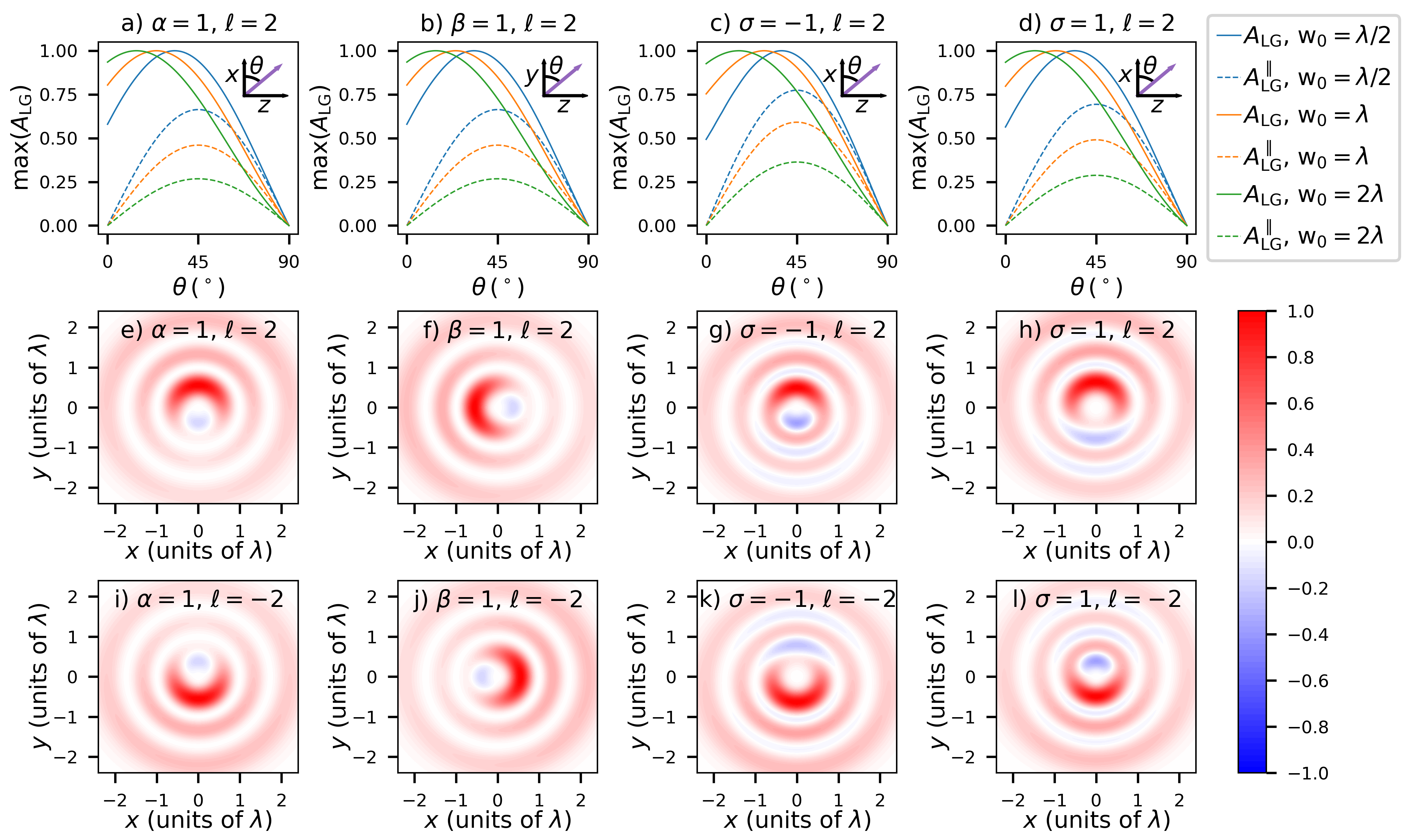}
    \caption{Same as Fig.~2 but for $p=2$ and $\ell=2$. }
    \label{fig:5}
\end{figure*}

Laguerre-Gaussian modes are described by both $\ell$ and the radial index $p$. We have observed how the sign of $\ell$ influences TD; here we study the role of the magnitude of $\ell$ and $p$. Fig.~\ref{fig:3} highlights the linear dependence on the magnitude of $\ell$ for TD as we see compared to the $\ell = 1$ case of Fig.~\ref{fig:2}, for $\ell = 2$ the relative contribution of the TD effect increases with increasing $\ell$. Figs.~\ref{fig:4} and \ref{fig:5} highlights the fact that increasing the radial order $p$ also yields larger relative TD contributions. This is because the magnitude of $p$ controls the $p+1$ concentric rings in the spatial distribution of Laguerre-Gaussian modes. Thus, increasing $p$ increases the transverse gradients of the beam, and within the paraxial approximation to first-order in $1/kw$ we can see from Eq.~(S1) in \cite{supplement} that increasing the transverse gradient increases the magnitude of the first-order longitudinal field component. The TD effect stems from the interference between the transverse and longitudinal field, and thus increasing $p$ increases $E_z$ which in turn makes the TD larger for increasing values of $p$. It is interesting to also see that increasing $p$ increases TD more significantly relative to increasing $\ell$. Finally, Figures~S4--S6 in \cite{supplement} display the results of the absorption by $\ell =0$ beams (i.e. non-vortex Gaussian). Interestingly, a tightly-focused 2D circularly-polarized beam does possess an in-phase relationship between transverse and longitudinal components, but the effect is significantly smaller than the effects due to the topological charge of vortex beams, and of course for 2D linear states there are no in-phase relationships between transverse and longitudinal components for non-vortex modes. 

The magnitude of TD with respect to the standard dichroic absorption mechanism increases with a tighter focus (smaller $w_0$); with the value of OAM through a larger $\ell$; using the anti-parallel combination of $\ell$ and $\sigma$; increasing the radial order $p$ and manipulating the ratio of ${\hat{z_i}\mu_i}/{\hat{x_j}\mu_j}$, which can be achieved by inherent material structure and/or orientation of the absorbing particle/structure. One-dimensional (1D) and 2D nanostructures are exemplary structures to exhibit $\ell$-dependent absorption, with the long-axis oriented parallel to the direction of beam propagation. This class of structure is already extremely well-utilized in applications in photonics and opto-electronics. Examples of these advanced materials which can exhibit TD include liquid crystals, carbon nanotubes and nanoribbons, metamaterials (e.g. hyperbolic plasmonic nanorods \cite{aigner2022nanophotonics}), etc. Furthermore, the diverse range of monolayer graphene derivatives and transition metal dichalcogenides \cite{chowdhury2020progress}, including van der Waals  heterostructures \cite{xiang2020one,guo2021one}, also constitute suitable materials to exhibit TD. Clearly, due to its local nature, TD lends itself to nano-optical probing methods. For example, TD will lead to a local loss of intensity and changing polarization azimuth in the planes orthogonal to \emph{z}. It therefore may be quantified by spatially resolved transmission and 3D Stokes polarimetry techniques. Alternatively, imaging of oriented fluorescent dipole emitters \cite{novotny2001longitudinal} (or photoluminescence in more general materials) would highlight this effect, and have the distinct advantage of a well-defined linearly oriented absorption dipole moment. 

\section{Topological-charge-dependent refraction} 

Alongside absorptive (dichroism) interactions, in transparent regions of materials there still manifests differential responses to the scattering of electromagnetic waves. One must be careful with the terminology here: scattering is more generally associated with extinction and a depleted transmission of the input light. However, both linear and circular birefringence at the microscopic level are in fact consequences of elastic scattering in the forward direction, i.e. coherent scattering (as is all refraction) \cite{barron2009molecular,jackson1999classical}. To elucidate the qualitative principle of topologically dependent refraction we use the simplest model of a `dilute' system (low number density of non-polar molecules, neglecting local field effects, etc.). The amplitude for elastic forward scattering, which produces the dynamic Stark shift $\Delta{E}$ and is also responsible for optical trapping \cite{andrews2016optical}, can be related to the refractive index of the molecular medium \cite{barron2009molecular,atkins2011molecular}. The forward scattering amplitude of Eq. \eqref{eq:1} for a number density \emph{N} of scattering centres is readily calculated using second-order perturbation theory (see \cite{supplement} for further information):

\begin{align}
\Delta{E}  &= \nonumber
-\frac{N I_\text{LG}(r,z)}{2c \epsilon_0}
\bigl[|\alpha|^2{\hat{x_i}\hat{x_j}} + |\beta|^2{\hat{y_i}\hat{y_j}}+ 2\Re\alpha\beta^*\hat{x_i}\hat{y_j}
\nonumber\\ 
&+2\hat{x_i}\hat{z_j}\frac{1}{k}\bigl\{(|\alpha|^2\frac{\ell}{r}-\gamma\Im\alpha^*\beta)\sin\phi \nonumber \\
&-\frac{\ell}{r}\Re\alpha\beta^*\cos\phi\bigr\} 
+2\hat{y_i}\hat{z_j}\frac{1}{k}\bigl\{(-\gamma\Im\alpha\beta^* \nonumber \\
&- \frac{\ell}{r}|\beta|^2) \cos\phi +\frac{\ell}{r}\Re\alpha\beta^*\sin\phi \bigr\}\bigr]\alpha_{ij}(\omega),
\tag{6}
\label{eq:6}
\end{align}

where $\alpha_{ij}(\omega)$ is the polarizability of the scattering centre and $I_\text{LG}(r,z)$ is the intensity of the LG beam. Accounting for energy conservation between the energy density of the field in vacuum and the energy shift experienced by the material, Eq. \eqref{eq:6} leads to the following refractive index (see \cite{supplement}): 

\begin{align}
n  &= \nonumber
1+\frac{N}{2 \epsilon_0}
\bigl[|\alpha|^2{\hat{x_i}\hat{x_j}} + |\beta|^2{\hat{y_i}\hat{y_j}}+ 2\Re\alpha\beta^*\hat{x_i}\hat{y_j}
\nonumber\\ 
&+2\hat{x_i}\hat{z_j}\frac{1}{k}\bigl\{(|\alpha|^2\frac{\ell}{r}-\gamma\Im\alpha^*\beta)\sin\phi \nonumber \\
&-\frac{\ell}{r}\Re\alpha\beta^*\cos\phi\bigr\} 
+2\hat{y_i}\hat{z_j}\frac{1}{k}\bigl\{(-\gamma\Im\alpha\beta^* \nonumber \\
&- \frac{\ell}{r}|\beta|^2) \cos\phi +\frac{\ell}{r}\Re\alpha\beta^*\sin\phi \bigr\}\bigr]\alpha_{ij}(\omega).
\tag{7}
\label{eq:7}
\end{align}

The refractive index Eq. \eqref{eq:7} clearly exhibits birefringence with respect to the sign of the topological charge, i.e. $n_{\ell} \neq n_{-\ell}$. A simple indicative calculation of the change in local refractive index due to topological charge birefringence using optimal conditions (see Fig. \ref{fig:2}a, $w_0=\lambda$) for 2D $x$-polarized light gives $\Delta n = n_{\ell} - n_{-\ell} = 0.6$, which is similar in magnitude to well-known linear birefringence of anisotropic crystals. The Stark shift Eq. \eqref{eq:6} exhibits the same spatial distribution as TD (see Fig. \ref{fig:2} -- \ref{fig:5}). Thus, in addition to TD, there also manifests $\ell$ dependent forward elastic scattering (topological-charge-dependent birefringence) of focused vortex beams at transparent frequencies in oriented materials. 

\section{Conclusion}

There has recently been a significant interest in light-matter interactions which are dependent on the wavefront handedness of an optical vortex through the sign of $\ell$ \cite{forbes2021orbital,green2023optical, forbes2023customized,porfirev2023light,andrews2023fundamental}. Such phenomena have primarily been concerned with chiral media and the optical chirality of vortex beams, manifesting through the interference of electric dipole coupling with the higher-order magnetic dipole and electric quadrupole interactions. These chiral effects are proportional to the second-order paraxial parameter $1/(kw)^2$ and require multipolar moments which are roughly 1000 times smaller than electric dipole coupling. Nonetheless, such effects have been experimentally observed \cite{wozniak2019interaction,ni2021gigantic,rouxel2022hard} and yield enhanced signals and sensitivity with respect to traditional plane-wave sources. Here we have highlighted absorption and forward elastic scattering (refraction) of light by oriented media which depends upon the sign of the topological charge of the input structured optical vortex beam through purely electric dipole interactions (i.e. not requiring chiral materials nor small multipolar couplings) and which are first-order in the paraxial parameter $1/(kw)$. The mechanisms we have discussed in this study should therefore be readily observable, indeed they are of the same order of paraxial parameter as the transverse SAM density of light which is a well-established experimental phenomenon \cite{eismann2021transverse,aiello2015transverse}. Compared to existing methods which exploit the handedness associated with the topological charge of vortex beams, techniques based on the phenomena described in this work should be more broadly applicable due to manifesting through purely electric-dipole couplings, thus representing potentially useful methods in the rapidly expanding toolkit of twisted light-matter interactions \cite{babiker2018atoms,forbes2021orbital,rosen2022interplay,andrews2023fundamental,porfirev2023light}. 

The mechanisms we have highlighted have their origins in the fundamental observation that focused optical vortices possess a polarization state which is dependent on the topological properties of the beam, specifically the value of $\ell$. As such, the topologically dependent absorption (topological-charge-dependent dichrosim, TD) and forward elastic scattering (topological-charge-dependent birefringence) are generic light-matter interactions for oriented media, and thus open the door for a whole array of specific applications in a wide range of advanced materials. Such quantitative studies represent a rich vein of future research. 

\section{Acknowledgements}

David L. Andrews is thanked for comments.

\bibliographystyle{apsrev4-1}
\bibliography{Main.bib}
\end{document}

\title{Supplementary Information: Topological-charge-dependent dichroism and birefringence of optical vortices}% Force line breaks with \\
\author{Kayn A. Forbes}
\email{K.Forbes@uea.ac.uk}

\affiliation{School of Chemistry, University of East Anglia, Norwich Research Park, Norwich NR4 7TJ, United Kingdom}

\author{Dale Green}
\affiliation{Physics, Faculty of Science, University of East Anglia, Norwich Research Park, Norwich NR4 7TJ, United Kingdom}

\maketitle

\section{Electric field of focused Laguerre-Gaussian beam}

The far-field paraxial electric field for a \emph{z}-propagating Laguerre-Gaussian beam has the following well-known form:

\begin{align}
\mathbf{E} &= \bigl[\alpha\mathbf{\hat{x}}
+\beta\mathbf{\hat{y}} 
\bigr]u_{\ell,p}^\text{LG}(r,\phi,z), %\tag{S1}
\label{eq:S1}
\end{align}
where $\alpha$ and $\beta$ are the Jones vectors describing the state of 2D polarization, and $|\alpha|^2 + |\beta|^2 = 1$. The amplitude distribution is given by the well-known formula (which also includes the time-dependence):

\begin{align}
u_{\ell,p}^\text{LG}(r,\phi,z) & = \nonumber \sqrt{\frac{2p!}{{\pi w_{0}^2}(p+\|\ell|)!}}\frac{w_0}{w[z]}
\Biggr(\frac{\sqrt{2}r}{w[z]}\Biggr)^{|\ell|}
\\ & \nonumber \times L_{p}^{|\ell|}\Biggr[\frac{2r^2}{w^2[z]}\Biggr] \text{exp}(-r^2/w^2[z])
\\ &  \nonumber  \times \text{exp}i(kz+\ell\phi+kr^2/2R[z] -\omega t
\\ & - (2p + |\ell| + 1)\zeta [z]). %\tag{S2}
\label{eq:S2}
\end{align}

Most quantities in Eq.~\eqref{eq:S2} are defined in the main manuscript; square brackets [] are reserved for arguments/dependence; $R[z]$ is the wavefront curvature; $\zeta[z]$ is the Gouy phase; and $L_{p}^{|\ell|}\bigr[\frac{2r^2}{w^2[z]}\bigr]$ is the generalized Laguerre polynomial. The first post-paraxial correction to Eq.~\eqref{eq:S1}, which represents the first-order longitudinal component, is easily calculated using Gauss's Law \cite{forbes2021relevance, adams2018optics}:

\begin{align}
E_z = \frac{i}{k}\Big(\alpha \pdv{}{x} + \beta \pdv{}{y}\Big)u_{\ell,p}^\text{LG}(r,\phi,z). %\tag{S3}
\label{eq:S3}
\end{align}

In using this formula the \emph{z}-dependence of the beam waist, wavefront curvature, and Gouy phase is ignored due to the constraint of working under a first-order approximation. Calculating Eq.~\eqref{eq:S3} using Eq.~\eqref{eq:S2} gives the first-order post paraxial electric field correction term, otherwise known as the (first-order) longitudinal field component. This longitudinal component when added to the transverse part Eq.~\eqref{eq:S1} gives the full first-order post paraxial electric field for a Laguerre-Gaussian beam, explicitly given in the main manuscript.

\section{Dynamic Stark effect}

\begin{figure} [h]
\includegraphics[width=1\linewidth]{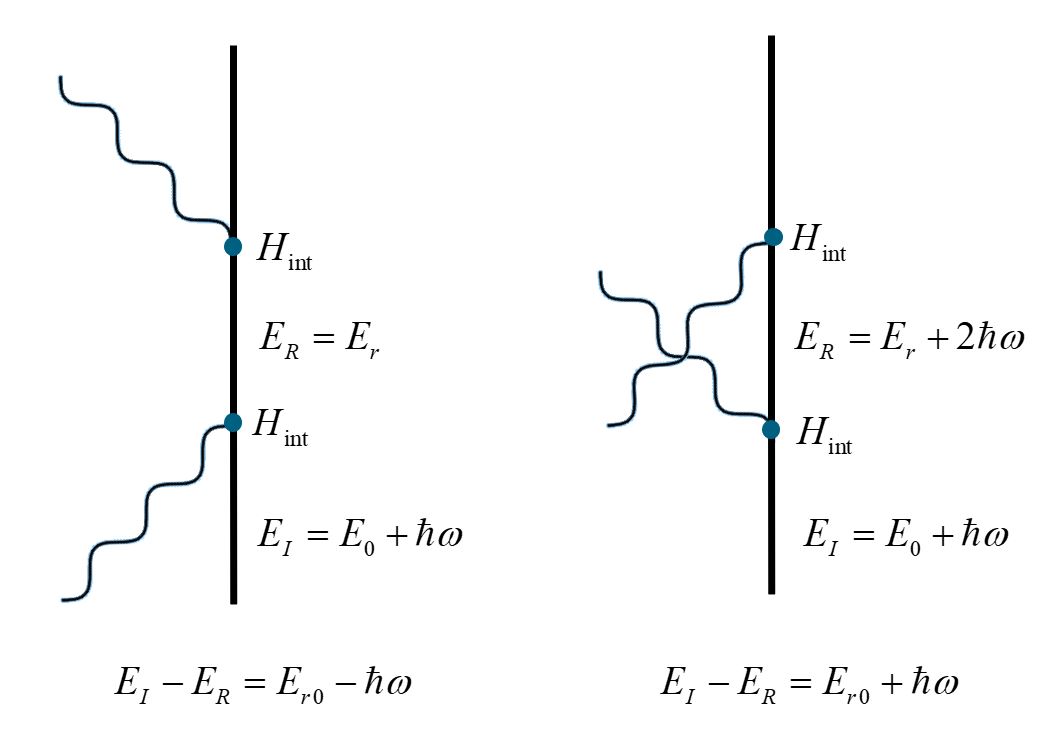}
\caption{The two Feynman diagrams required in the calculation of single-photon scattering. The notation $E_{r0}$ is shorthand for $E_{r0} = E_r-E_0$.} 
    \label{fig:S1}
\end{figure}

The dynamic stark shift $\Delta{E}$ is calculated using second-order perturbation theory \cite{craig1998molecular, andrews2016optical}:

\begin{align}
\Delta{E}  = \Re \sum_R \frac{\langle I | H_\text{int} | R \rangle \langle R | H_\text{int} | I \rangle }{E_I-E_R}, 
%\tag{S4}
\label{eq:S4}
\end{align}
where $R$ denotes short-lived virtual states. Compared to semi-classical methods, the explicit calculation of Eq.~\eqref{eq:S4} is exceedingly straightforward using non-relativistic quantum electrodynamics \cite{andrews2016optical, craig1998molecular}. We therefore take this approach for simplicity. In order to do so, however, we need to use the quantized mode expansion for the electric field (Eq 1. from the main manuscript):

\begin{align}
\hat{\mathbf{E}} &= \sum_{k,\eta,\ell,p}\Omega \Bigl(\bigl[\alpha\mathbf{\hat{x}}
+\beta\mathbf{\hat{y}} 
+\frac{i}{k}\mathbf{\hat{z}}
\bigl\{\alpha\bigl(\gamma\cos\phi-\frac{i\ell}{r}\sin\phi\bigr)
\nonumber\\ &+\beta\bigl(\gamma\sin\phi+\frac{i\ell}{r}\cos\phi\bigr)\bigr\} \bigr]\hat{a}\Tilde{u}_{\ell,p}^{\text{LG}}-\text{H.c.}\Bigr),
%\tag{S5}
\label{eq:S5}
\end{align}
where the normalization constant is given as $\Omega = i(\hbar ck/2 \epsilon_0 A_{\ell,p} V)^\frac{1}{2}$; $\hat{a}$ is lowering operator, its Hermitian conjugate (H.c.) is the raising operator $\hat{a}^\dagger$; $\eta$ denotes the 2D polarization state; $A_{\ell,p}$ is a normalization constant for LG beams, it has dimensions of $\text{length}^2$; $V$ is the quantization volume;  and $\Tilde{u}_{\ell,p}^{\text{LG}}$ is Eq.~\eqref{eq:S2} without the time-dependent exponential. For coherent forward Rayleigh scattering of a beam of photons in a mode $(k,\eta,\ell,p)$, $| I \rangle$ = $| E_0; n(k,\eta,\ell,p) \rangle$. The virtual intermediate states $| R \rangle$ of the light-matter system can be calculated with the aid of two topologically distinct time-ordered Feynman diagrams in Fig.~\ref{fig:S1}. Using the dipole interaction Hamiltonian $H_\text{int} = -\hat{\mu}_i \hat{E}_i$ (applicable to a single particle) and Eq.~\eqref{eq:S5} in Eq.~\eqref{eq:S4} yields

\begin{align}
\Delta{E}  &=  n \Omega^2 |\Tilde{u}_{\ell,p}^{\text{LG}}|^2 
\text{Re} \bigl[\alpha{\hat{x}_i}
+\beta{\hat{y}_i} +\frac{i}{k}{\hat{z}_i}
\bigl\{\alpha\bigl(\gamma\cos\phi
\nonumber\\
&-\frac{i\ell}{r}\sin\phi\bigr)+\beta\bigl(\gamma\sin\phi+\frac{i\ell}{r}\cos\phi\bigr)\bigr\} \bigr] 
\nonumber\\
&  \times \bigl[{\alpha}^*{\hat{x}_j}
+\beta^*{\hat{y}_j}  -\frac{i}{k}{\hat{z}_j}\bigl\{\alpha^*\bigl(\gamma\cos\phi+\frac{i\ell}{r}\sin\phi\bigr) 
\nonumber\\
& +\beta^*\bigl(\gamma\sin\phi-\frac{i\ell}{r}\cos\phi\bigr)\bigr\} \bigr] \nonumber\\
&
\times \sum_r \Bigg[\frac{\mu{_i}^{0r}\mu{_j}^{r0}}{E_{r0}-\hbar\omega}+\frac{\mu{_j}^{0r}\mu{_i}^{r0}}{E_{r0}+\hbar\omega} \Bigg].
%\tag{S6}
\label{eq:s6}
\end{align}

In deriving the above we assume a high photon number $n$ such that the photon number coming from right hand graph in Fig.~\ref{fig:S1} $n+1$ is approximated as $n$. The quantity in the final line of Eq.~\eqref{eq:s6} is the well-known frequency-dependent (dynamic) polarizability tensor $\alpha_{ij}^{00}(\omega,-\omega)$.
The intensity of the beam is given as

\begin{align}
I_\text{LG}(r,z) &= \frac{n \hbar c^2 k }{A_{\ell,p} V}|\Tilde{u}_{\ell,p}^{\text{LG}}|^2, 
%\tag{S7}
\label{eq:s7}
\end{align}

which allows Eq.~\eqref{eq:s6} to be written as 

\begin{align}
\Delta{E}  &=
-\frac{I_\text{LG}(r,z)}{2c \epsilon_0}\bigl[|\alpha|^2{\hat{x_i}\hat{x_j}} + |\beta|^2{\hat{y_i}\hat{y_j}}+ 2\Re\alpha\beta^*\hat{x_i}\hat{y_j}
\nonumber\\ 
&+2\hat{x_i}\hat{z_j}\frac{1}{k}\bigl\{(|\alpha|^2\frac{\ell}{r}-\gamma\Im\alpha^*\beta)\sin\phi \nonumber \\
&-\frac{\ell}{r}\Re\alpha\beta^*\cos\phi\bigr\} 
+2\hat{y_i}\hat{z_j}\frac{1}{k}\bigl\{(-\gamma\Im\alpha\beta^* \nonumber \\
&- \frac{\ell}{r}|\beta|^2) \cos\phi +\frac{\ell}{r}\Re\alpha\beta^*\sin\phi \bigr\}\bigr]
\alpha_{ij}^{00}(\omega,-\omega).
%\tag{S8}
\label{eq:S8}
\end{align}

\section{Topological-charge-dependent refraction}

The cycle-averaged energy density of a Laguerre-Gaussian beam in vacuum (restricting calculations to the paraxial approximation mentioned in the main article) is 

\begin{align}
\overline{W}{_{\text{E}}^{\text{vac}}} &= \nonumber \frac{\epsilon_0}{2}\text{Re}\bigl(\mathbf{E^*}\cdot{\mathbf{E}}\bigr)
\\&=\frac{\epsilon_0}{2}|u_{\ell,p}^{\text{LG}}|^2.
%\tag{S9}
\label{eq:S9}
\end{align}

The energy density of this beam in a medium of refractive index $n$ is thus $\overline{W}{_{\text{E}}^{\text{med}}}=\frac{n\epsilon_0}{2}|u_{\ell,p}^{\text{LG}}|^2$. Under off-resonant laser light the medium experiences the dynamic Stark Effect  which lowers the energy of the medium: 

\begin{align}
\Delta{E}  &= \nonumber
-\frac{N I_\text{LG}(r,z)}{2c \epsilon_0}
\bigl[|\alpha|^2{\hat{x_i}\hat{x_j}} + |\beta|^2{\hat{y_i}\hat{y_j}}+ 2\Re\alpha\beta^*\hat{x_i}\hat{y_j}
\nonumber\\ 
&+2\hat{x_i}\hat{z_j}\frac{1}{k}\bigl\{(|\alpha|^2\frac{\ell}{r}-\gamma\Im\alpha^*\beta)\sin\phi \nonumber \\
&-\frac{\ell}{r}\Re\alpha\beta^*\cos\phi\bigr\} 
+2\hat{y_i}\hat{z_j}\frac{1}{k}\bigl\{(-\gamma\Im\alpha\beta^* \nonumber \\
&- \frac{\ell}{r}|\beta|^2) \cos\phi +\frac{\ell}{r}\Re\alpha\beta^*\sin\phi \bigr\}\bigr]\alpha_{ij}(\omega),
%\tag{S10}
\label{eq:S10}
\end{align}

where $N$ is the number density of scattering particles. Eq.~\eqref{eq:S10} is of course just Eq.~\eqref{eq:S8} multiplied by $N$. Due to energy conservation $\Delta{E}  =\overline{W}{_{\text{E}}^{\text{vac}}}-\overline{W}{_{\text{E}}^{\text{med}}}=\overline{W}{_{\text{E}}^{\text{vac}}}\bigl(1-n\bigr)$, and therefore $n=1-\frac{\Delta{E}}{\overline{W}{_{\text{E}}^{\text{vac}}}}$:

\begin{align}
n  &= \nonumber
1+\frac{N}{2 \epsilon_0}
\bigl[|\alpha|^2{\hat{x_i}\hat{x_j}} + |\beta|^2{\hat{y_i}\hat{y_j}}+ 2\Re\alpha\beta^*\hat{x_i}\hat{y_j}
\nonumber\\ 
&+2\hat{x_i}\hat{z_j}\frac{1}{k}\bigl\{(|\alpha|^2\frac{\ell}{r}-\gamma\Im\alpha^*\beta)\sin\phi \nonumber \\
&-\frac{\ell}{r}\Re\alpha\beta^*\cos\phi\bigr\} 
+2\hat{y_i}\hat{z_j}\frac{1}{k}\bigl\{(-\gamma\Im\alpha\beta^* \nonumber \\
&- \frac{\ell}{r}|\beta|^2) \cos\phi +\frac{\ell}{r}\Re\alpha\beta^*\sin\phi \bigr\}\bigr]\alpha_{ij}(\omega).
%\tag{S11}
\label{eq:S11}
\end{align}

\section{Additional plots of the spatial variation and dipole orientation dependence of topological-charge-dependent dichroism}

\begin{figure*}
    \includegraphics[]{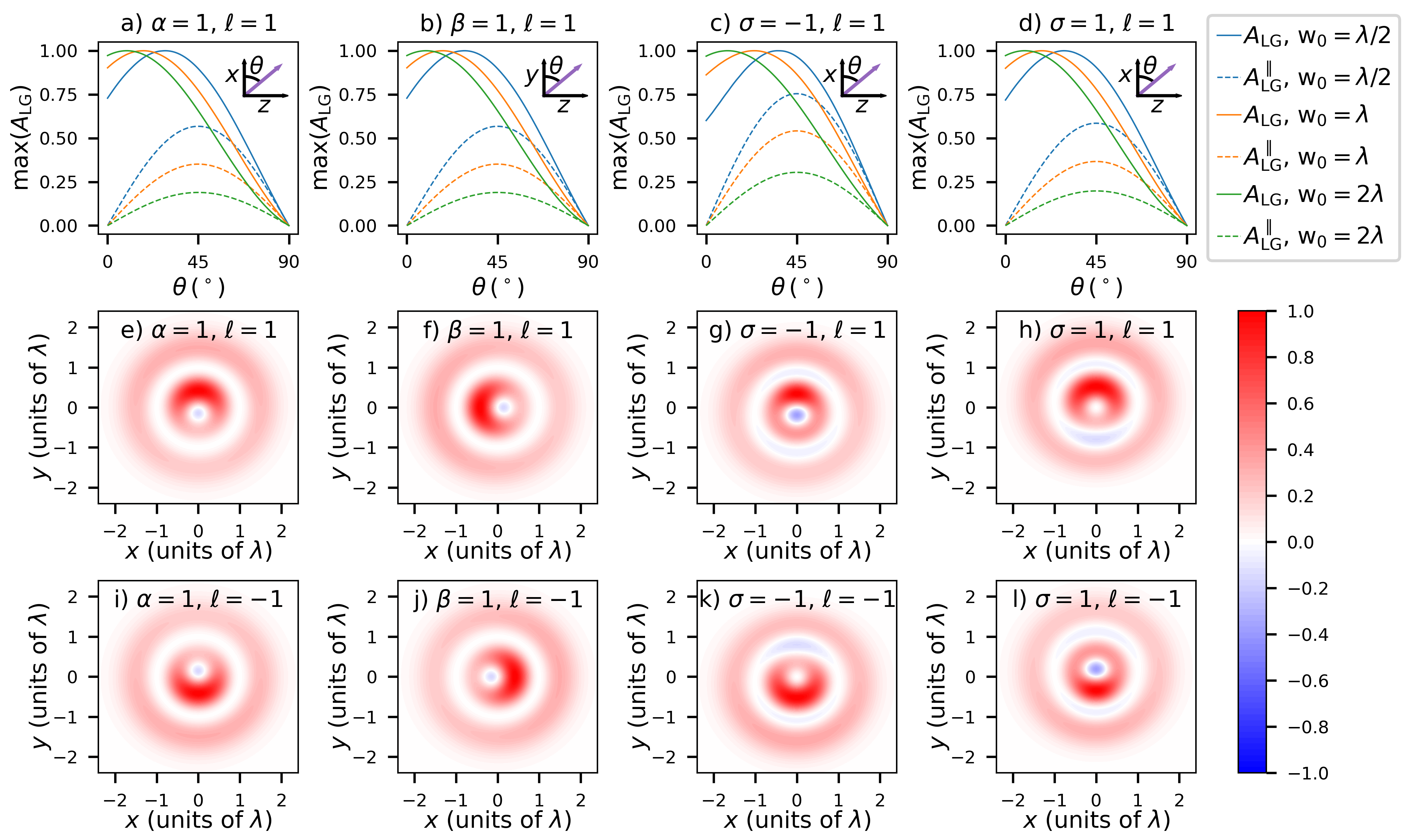}
    \caption{Same as Fig.~2 of the main manuscript but for $p=1$.}
    \label{fig:S3}
\end{figure*}

\begin{figure*}
    \includegraphics[]{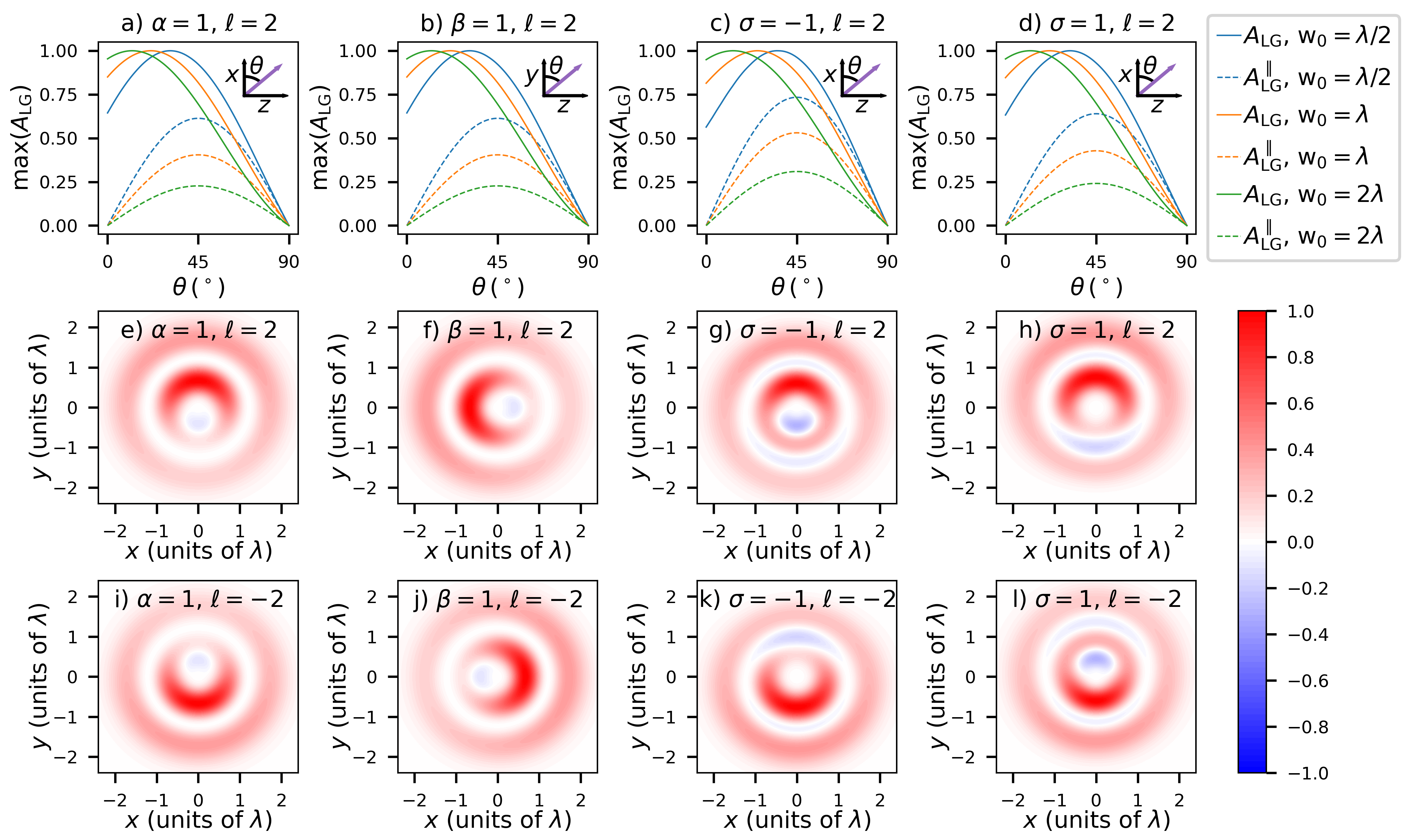}
    \caption{Same as Fig.~2 of the main manuscript but for $p=1$ and $\ell=2$.}
    \label{fig:S4}
\end{figure*}

\begin{figure*}
    \includegraphics[]{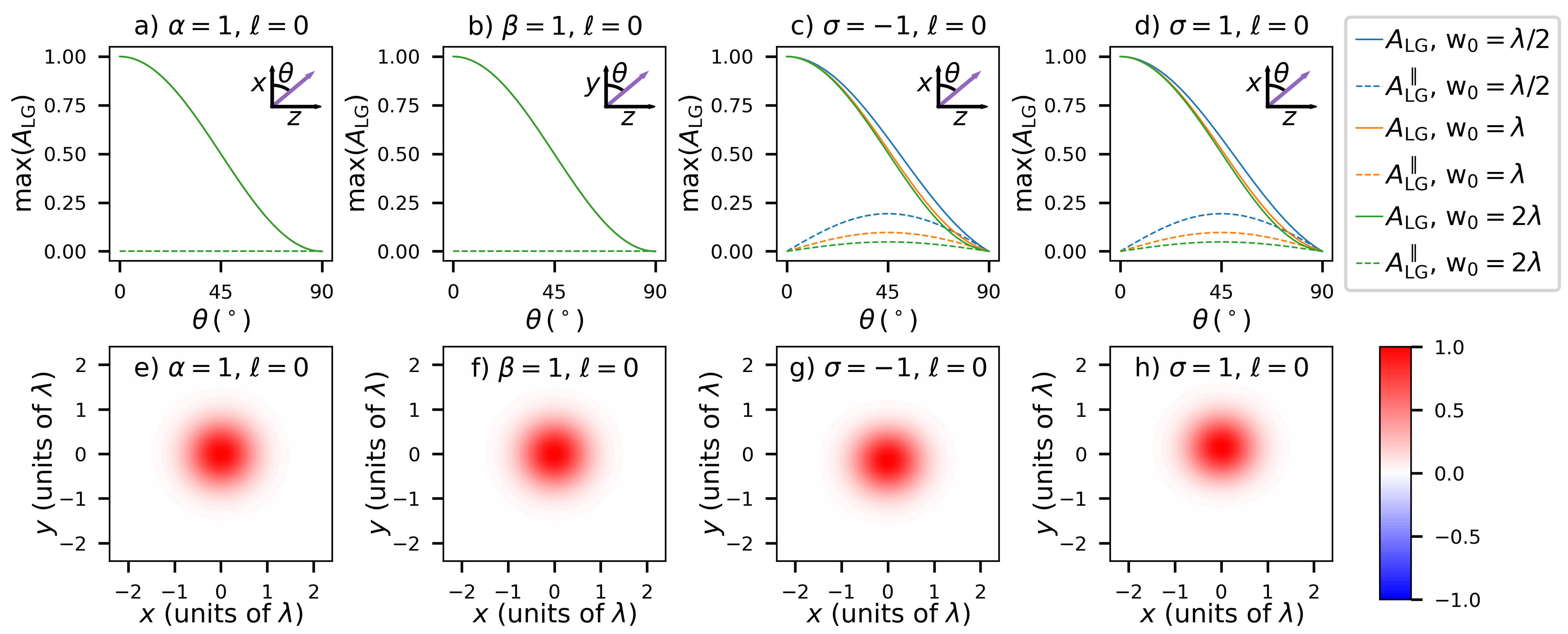}
    \caption{Same as Fig.~2 of the main manuscript but for $p=0$ and $\ell=0$, i.e. a Gaussian beam.}
    \label{fig:S7}
\end{figure*}

\begin{figure*}
    \includegraphics[]{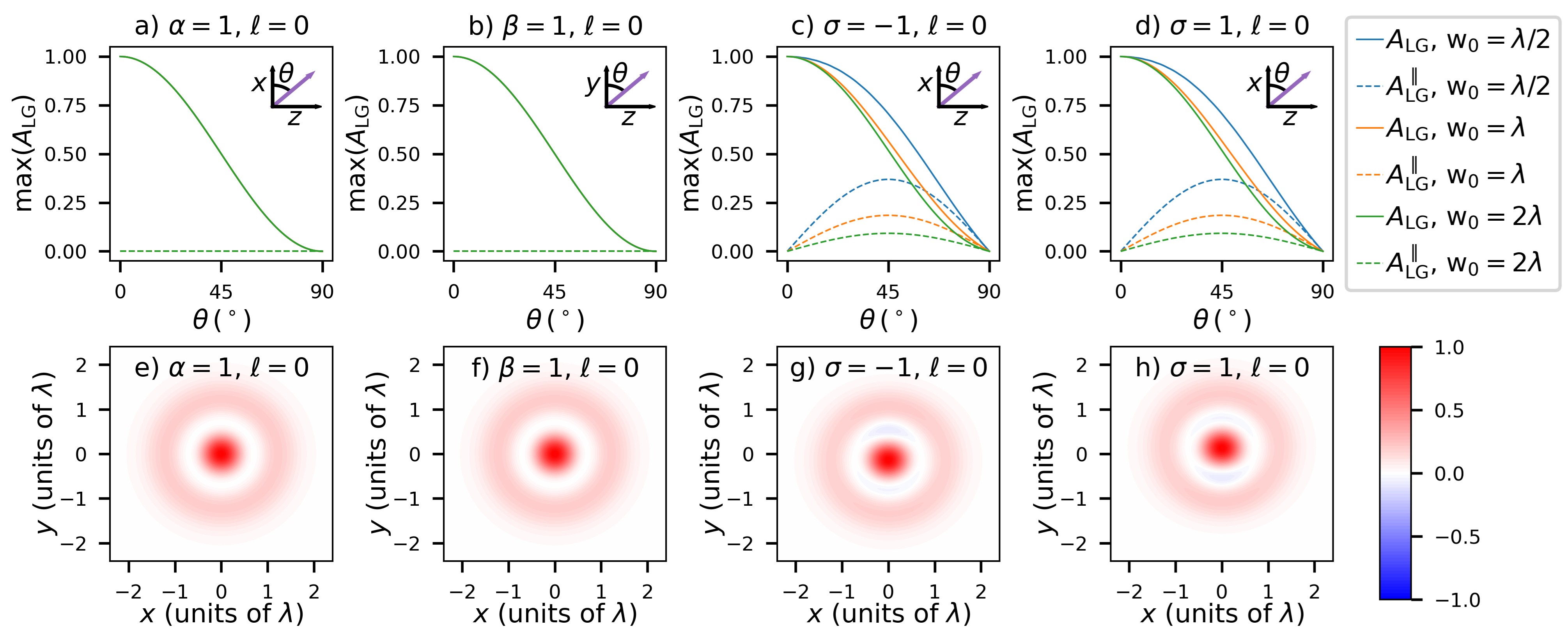}
    \caption{Same as Fig.~2 of the main manuscript but for $p=1$ and $\ell=0$.}
    \label{fig:S8}
\end{figure*}

\begin{figure*}
    \includegraphics[]{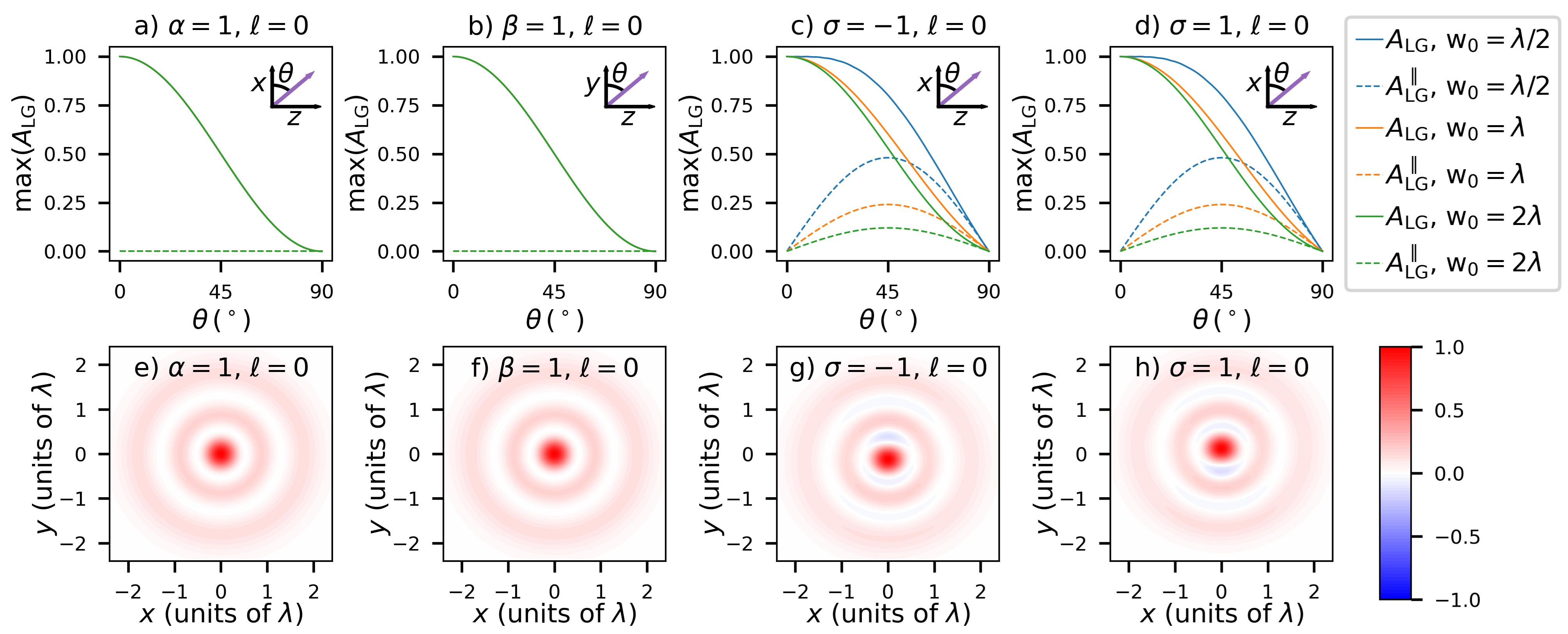}
    \caption{Same as Fig.~2 of the main manuscript but for $p=2$ and $\ell=0$. }
    \label{fig:S9}
\end{figure*}

\bibliographystyle{apsrev4-1}
\bibliography{Main.bib}